\documentclass[article,12pt] {revtex4}
\usepackage{amsfonts}
\usepackage{graphicx, amsmath}
\begin{document}

\title[Short Title]{Generation of three-dimensional entanglement between two spatially separated atoms via shortcuts to adiabatic passage}
\author{Jing-bo Lin}
\affiliation{Department of Physics, College of Science, Yanbian
University, Yanji, Jilin 133002, People's Republic of China}
\author{Yan Liang}
\affiliation{Department of Physics, College of Science, Yanbian
University, Yanji, Jilin 133002, People's Republic of China}
\author{Xin Ji \footnote{E-mail: jixin@ybu.edu.cn}}
\affiliation{Department of Physics, College of Science, Yanbian University, Yanji, Jilin
133002, People's Republic of China}
\author{Shou Zhang}
\affiliation{Department of Physics, College of Science, Yanbian University, Yanji, Jilin
133002, People's Republic of China}
\begin{abstract}

\noindent {\bf Abstract}

We propose a scheme for generating three-dimensional entanglement
between two atoms trapped in two spatially separated cavities
reapectively via shortcuts to adiabatic passage based on the
approach of Lewis-Riesenfeld invariants in cavity quantum
electronic dynamics. By combining Lewis-Riesenfeld invariants with
quantum Zeno dynamics, we can generate three dimensional
entanglement of the two atoms with high fidelity. The Numerical
simulation results show that the scheme is robust against the
decoherences caused by the photon leakage and atomic spontaneous
emission.

\end{abstract}

\maketitle
\noindent {\bf 1. Introduction}

Quantum entanglement plays a significant role not only in testing
quantum nonlocality, but also in a variety of quantum information
tasks
\cite{AKE,CHB,CHBS,CHBG,KMH,MHVB,SBZGCG,GV,SHT1997,MI2000,CD2000,JWDD1996}.
Recently, high-dimensional entanglement is becoming more and more
important since they are more secure than qubit systems,
especially in the aspect of quantum key distribution. Besides, it
has been demonstrated that violations of local realism by two
entangled high-dimensional systems are stronger than that by
two-dimensional systems \cite{DPMW2000}. So a lot of works have
been done theoretically and experimentally in generating
high-dimensional entanglement
\cite{WG2011,LP2012,WCYC2013,QCW2013,XQ201405,XQ201402,SL2014,YLS2015,AAGA2001,AGA2002}.

In order to realize the entanglement generation or population
transfer in a quantum system with time-dependent interacting
field, many schemes have been put forward. Such as $\pi$ pulses,
composite pulses, rapid  adiabatic passage(RAP), stimulated Raman
adiabatic passage , and their variants
\cite{KHB1998,PIMS2007,NVTB2001}. STIRAP is widely used in
time-dependent interacting field because of the robustness for
variations in the experimental parameters. But it usually requires
a relatively long interaction time, so that the decoherence would
destroy the intended dynamics, and finally lead to an error
result. Therefore, reducing the time of dynamics towards the
perfect final outcome is necessary and perhaps the most effective
method to essentially fight against the dissipation which comes
from noise or losses accumulated during the operational processes.
Rencently, various schemes have been explored theoretically and
experimentally to construct shortcuts for adiabatic passage
\cite{XASA2010,KPYR2011,JXPP2011,ARXD2012,AFTS2012,AC2013,MYLJ2014,YHC2014,YLQ2015,YLC2015,YLX2015}.
Unfortunately, as far as we know, the research of constructing
shortcuts to adiabatic passage for generating entanglement has not
been comprehensively studied.

In this paper, we construct an effective shortcuts to adiabatic
passage for generating three dimentional entanglement between two
atoms trapped in two spatially separated cavities connected by a
fiber based on the Lewis-Riesenfeld invariants and quantum Zeno
dynamics (QZD). The time for generating three dimentional
entanglement in our scheme is much shorter time than that based on
adiabatic passage technique. Moreover, the strict numerical
simulations demonstrate that our scheme is insensitive to the
decoherence caused by spontaneous emission and photon leakage.

This paper is structured as follows: In Section 2, we give a brief
description about Lewis-Riesenfeld invariants and QZD. In Section
3, we construct a shortcuts for generating three dimentional
entanglement. Section 4 shows the numerical simulation results and
feasibility analysis. The conclusion appears in Section 5.

\noindent {\bf 2. Preliminary theory}

\noindent{\bf 2.1. Lewis-Riesenfeld invariants}

We first give a brief description about Lewis-Riesenfeld
invariants theory \cite{HRL1969,MAL2009}. A quantum system is
governed by a time-dependent Hamiltonian $H(t)$, and the
corresponding time-dependent Hermitian invariant $I(t)$ satisfies
\begin{eqnarray}\label{1}
i\hbar \frac{\partial I(t)}{\partial t}&=&[H(t),I(t)].
\end{eqnarray}
The solution of the time-dependent Schr\"odinger equation $i\hbar$
$\frac{\partial |\Psi(t)\rangle}{\partial t}$ $
=H(t)|\Psi(t)\rangle$ can be expressed by a superposition of
invariant $I(t)$ dynamical modes $|\Phi_{n}(t)\rangle$
\begin{eqnarray}\label{2}
|\Psi(t)\rangle&=&\sum_n C_n e^{i\alpha_n}|\Phi_{n}(t)\rangle,
\end{eqnarray}
where $C_n$ is time-independent amplitude, $\alpha_n$ is the
Lewis-Riesenfeld phase, $|\Phi_{n}(t)\rangle$ is one of the
orthogonal eigenvectors of the invariant $I(t)$, satisfying
$I(t)|\Phi_{n}(t)\rangle=\lambda_n|\Phi_{n}(t)\rangle$, with
$\lambda_n$ being real constant. And the Lewis-Riesenfeld phases
are defined as
\begin{eqnarray}\label{3}
\alpha_n(t)&=&\frac{1}{\hbar}\int_0^t dt^\prime
\langle\Phi_n(t^\prime )|i\hbar\frac{\partial}{\partial t^\prime
}-H(t^\prime )|\Phi_n(t^\prime )\rangle.
\end{eqnarray}

\noindent{\bf 2.2. Quantum Zeno dynamics}

Quantum Zeno effect is an interesting phenomenon in quantum
mechanics. Recent studies \cite{PVGS2000,PS2002,PGS2009} show that
a quantum Zeno evolution will evolve away from its initial state,
but it remains in the Zeno subspace defined by the measurements
\cite{PVGS2000} via frequently projecting onto a multidimensional
subspace. This is known as QZD. We consider a system which is
governed by the Hamiltonian
\begin{eqnarray}\label{4}
H_K=H_{\rm obs}+KH_{\rm meas},
\end{eqnarray}
where $H_{\rm obs}$ is the Hamiltonian of the investigated quantum
system and the $H_{\rm meas}$ is the interaction Hamiltonian
performing the measurement. $K$ is a coupling constant, and when
it satisfies $K\rightarrow \infty$, the whole system is governed
by the evolution operator
\begin{eqnarray}\label{5}
U(t)={\rm exp}[-it\sum_{n}(K\lambda_nP_n+P_nH_{\rm obs}P_n)],
\end{eqnarray}
where $P_n$ is one of the eigenprojections of $H_{\rm meas}$ with
eigenvalues $\lambda_n$($H_{\rm meas} = \sum_{n}\lambda_nP_n$).

\noindent {\bf 3. Shortcuts to adiabatic passage for generating
three-dimensional entanglement of two atoms}

\begin{figure}[ht]\centering
\scalebox{0.9}{\includegraphics{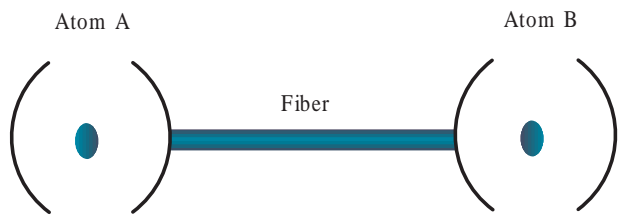}}\caption{The schematic
setup for generating two atoms three-dimensional entanglement. The
two atoms are trapped in two spatially separated optical cavities
connected by a fiber.}\label{fig1}
\end{figure}
\begin{figure}[ht]\centering
\scalebox{0.9}{\includegraphics{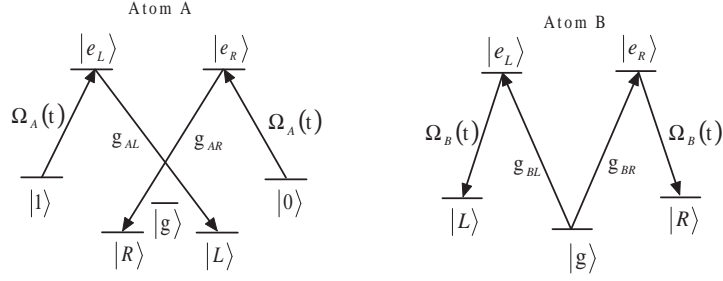}}\caption{The level
configurations of atom A and B.}\label{fig2}
\end{figure}

The schematic setup for generating three-dimensional entanglement
of two atoms is shown in Fig.\ref{fig1}. We consider a
cavity-fibre-cavity system, in which two atoms are trapped in the
corresponding optical cavities connected by a fiber. Under the
short fiber limit $(lv)/(2\pi c)\ll 1$, only the resonant mode of
the fiber will interact with the cavity mode \cite{SMB2006}, where
$l$ is the length of the fiber and $v$ is the decay rate of the
cavity field into a continuum of fiber modes. The corresponding
level structures of atoms are shown in Fig. \ref{fig2}. Atom A has
two excited states $|e_{L}\rangle,|e_{R}\rangle$, and five ground
states $|1\rangle$, $|R\rangle$, $|L\rangle$, $|g\rangle$ and
$|0\rangle$, while atom B is a five-level system with three ground
states $|R\rangle$, $|L\rangle$ and $|g\rangle$, two excited
states $|e\rangle_{L}$ and $|e\rangle_{R}$. For atom A, the
transitions $|0\rangle\leftrightarrow|e_{R}\rangle$ and
$|1\rangle\leftrightarrow|e_{L}\rangle$ are driven by classical
fields with the same Rabi frequency $\Omega_A(t)$. And the
transitions $|R\rangle\leftrightarrow|e_{R}\rangle$ and
$|L\rangle\leftrightarrow|e_{L}\rangle$ are resonantly driven by
the corresponding cavity mode $a_{Aj}$ with $j$-circular
polarization and the coupling strength is $g_{Aj}$ $(j=L,R)$. For
atom B, the transitions $|R\rangle\leftrightarrow|e_{R}\rangle$
and $|L\rangle\leftrightarrow|e_{L}\rangle$ are driven by
classical fields with the same Rabi frequency $\Omega_B(t)$, and
the transitions $|g\rangle\leftrightarrow|e_{R}\rangle$ and
$|g\rangle\leftrightarrow|e_{L}\rangle$ are resonantly driven by
the corresponding cavity mode $a_{Bj}$ with $j$-circular
polarization and the coupling strength is $g_{Bj}$ $(j=L,R)$. The
whole Hamiltonian in the interaction picture can be written as
($\hbar=1$):
\begin{eqnarray}\label{6}
H_{1}&=&H_{a\text{-}l}+H_{a\text{-}c\text{-}f},
\end{eqnarray}
\begin{eqnarray}\label{7}
H_{a\text{-}l}&=&\Omega_{A}(t)(|e_L\rangle_{A}\langle{1}|+|e_R\rangle_{A}\langle{0}|)+\Omega_{B}(t)(|e_L\rangle_{B}\langle{L}|
+|e_R\rangle_{B}\langle{R}|)+\rm H.c.,
\end{eqnarray}
\begin{eqnarray}\label{8}
H_{a\text{-}c\text{-}f}&=&g_{AL}a_{AL}|e_L\rangle_{A}\langle{L}|+g_{AR}a_{AR}|e_R\rangle_{A}\langle{R}|
+g_{BL}a_{BL}|e_L\rangle_{B}\langle{g}|+g_{BR}a_{BR}|e_R\rangle_{B}\langle{g}|\cr
&&+\eta b_{L}(a_{AL}^{\dag}+a_{BL}^{\dag})+\eta
b_{R}(a_{AR}^{\dag}+a_{BR}^{\dag})+\rm H.c.,
\end{eqnarray}
where $\eta$ is the coupling strength between cavity mode and the
fiber mode, $b_{R(L)}$ is the annihilation operator for the fiber
mode with $R(L)$-circular polarization, $a_{A(B)R(L)}$ is the
annihilation operator for the corresponding cavity field with
$R(L)$-circular polarization, and $g_{A(B)R(L)}$ is the coupling
strength between the corresponding cavity mode and the trapped
atom.

In order to obtain the following two atoms three-dimensional
entanglement:
\begin{eqnarray}\label{9}
|\Psi\rangle&=&\frac{1}{\sqrt{3}}(|R\rangle_A|R\rangle_B+|L\rangle_A|L\rangle_B+|g\rangle_A|g\rangle_B),
\end{eqnarray}
we assume atom A in the state
\begin{eqnarray}\label{10}
|\Psi_A\rangle&=&\frac{1}{\sqrt{3}}(|1\rangle_A+|0\rangle_A+|g\rangle_A),
\end{eqnarray}
while atom B in the state $|g\rangle_B$, both the cavity modes and
the fiber mode in vacuum state
$|0\rangle_{AC}|0\rangle_{BC}|0\rangle_f$. Then we present how to
realize the evolutions of the atom state $|1\rangle_A|g\rangle_B$
to $-|L\rangle_A|L\rangle_B$, $|0\rangle_A|g\rangle_B$ to
$-|R\rangle_A|R\rangle_B$, $|g\rangle_A|g\rangle_B$ to
$-|g\rangle_A|g\rangle_B$.

For the initial state
$|0\rangle_A|g\rangle_B|0\rangle_{AC}|0\rangle_{BC}|0\rangle_f$,
the whole system evolves in the subspace spanned by
\begin{eqnarray}\label{11}
|\phi_1\rangle&=&|0\rangle_A|g\rangle_B|0\rangle_{AC}|0\rangle_{BC}|0\rangle_f,\nonumber\\
|\phi_2\rangle&=&|e_R\rangle_A|g\rangle_B|0\rangle_{AC}|0\rangle_{BC}|0\rangle_f,\nonumber\\
|\phi_3\rangle&=&|R\rangle_A|g\rangle_B|1_R\rangle_{AC}|0\rangle_{BC}|0\rangle_f,\nonumber\\
|\phi_4\rangle&=&|R\rangle_A|g\rangle_B|0\rangle_{AC}|0\rangle_{BC}|1_R\rangle_f,\nonumber\\
|\phi_5\rangle&=&|R\rangle_A|g\rangle_B|0\rangle_{AC}|1_R\rangle_{BC}|0\rangle_f,\nonumber\\
|\phi_6\rangle&=&|R\rangle_A|e_R\rangle_B|0\rangle_{AC}|0\rangle_{BC}|0\rangle_f,\nonumber\\
|\phi_7\rangle&=&|R\rangle_A|R\rangle_B|0\rangle_{AC}|0\rangle_{BC}|0\rangle_f.
\end{eqnarray}
Seting $\Omega_{A}(t),\Omega_{B}(t)\ll \eta,g_{AR(L)}, g_{BR(L)}$,
then both the condition $ H_{a\text{-}c\text{-}f}\gg
H_{a\text{-}l}$ and the Zeno condition $ K \rightarrow \infty$ can
be satisfied ($ H_{a\text{-}l}$ and $ H_{a\text{-}c\text{-}f}$
correspond respectively to $ H_{obs}$ and $ {KH}_{meas}$ in Eq.
(\ref{4})). By performing the unitary transformation $U
=e^{-i\emph{H}_{\emph{a}\text{-}\emph{c}\text{-}\emph{f}}~\emph{t}}$
under condition $ H_{a\text{-}c\text{-}f}\gg H_{a\text{-}l}$, the
Hilbert subspace can be divided into five invariant Zeno subspaces
\cite{PS2002,PGS2009}:
\begin{eqnarray}\label{12}
\Gamma_{P1}&=&\Big\{|\phi_1\rangle,|\phi_7\rangle,|\psi_1\rangle\Big\},\nonumber\\
\Gamma_{P2}&=&\Big\{|\psi_2\rangle\Big\},~~~~~~~~~\Gamma_{P3}~=~\Big\{|\psi_3\rangle\Big\},\nonumber\\
\Gamma_{P4}&=&\Big\{|\psi_4\rangle\Big\},~~~~~~~~~\Gamma_{P5}~=~\Big\{|\psi_5\rangle\Big\},
\end{eqnarray}
with the eigenvalues $\lambda_1=0$, $\lambda_2=-g$, $\lambda_3=g$,
$\lambda_4=-\sqrt{g^2+2\eta^2}=-\varepsilon$, and
$\lambda_5=\sqrt{g^2+2\eta^2}=\varepsilon$, where we assume
$g_{AR(L)}=g_{BR(L)}=g$ for simplicity. Here
\begin{eqnarray}\label{13}
|\psi_1\rangle&=&\frac{1}{\varepsilon}(\eta|\phi_2\rangle-g|\phi_4\rangle+\eta|\phi_6\rangle),\nonumber\\
|\psi_2\rangle&=&\frac{1}{2}(-|\phi_2\rangle+|\phi_3\rangle-|\phi_5\rangle+|\phi_6\rangle),\nonumber\\
|\psi_3\rangle&=&\frac{1}{2}(-|\phi_2\rangle-|\phi_3\rangle+|\phi_5\rangle+|\phi_6\rangle),\nonumber\\
|\psi_4\rangle&=&\frac{1}{2\varepsilon}(g|\phi_2\rangle-\varepsilon|\phi_3\rangle+2\eta|\phi_4\rangle)-\varepsilon|\phi_5\rangle+g|\phi_6\rangle,\nonumber\\
|\psi_5\rangle&=&\frac{1}{2\varepsilon}(g|\phi_2\rangle+\varepsilon|\phi_3\rangle+2\eta|\phi_4\rangle)+\varepsilon|\phi_5\rangle+g|\phi_6\rangle,
\end{eqnarray}
and the corresponding projection
\begin{eqnarray}\label{14}
P_i^\alpha =
|\alpha\rangle\left\langle\alpha\right|,(|\alpha\rangle\in\Gamma_{Pi}).
\end{eqnarray}
Under the above condition, the system Hamiltonian can be rewritten
as the following form \cite{PGS2009}:
\begin{eqnarray}\label{15}
H_{\rm total}&\simeq&
\sum_{i,\alpha,\beta}(\lambda_iP_i^\alpha+P_i^\alpha H_{a\text{-}l} P_i^\beta) \nonumber\\
&=&-g|\psi_2\rangle\left\langle\psi_2\right|+g|\psi_3\rangle\left\langle\psi_3\right|-\varepsilon|\psi_4\rangle\left\langle\psi_4\right|+
\varepsilon|\psi_5\rangle\left\langle\psi_5\right|\cr
&&+\frac{1}{\varepsilon}\eta(\Omega_{A}(t)|\psi_1\rangle\left\langle\phi_1\right|+\Omega_{B}(t)|\psi_1\rangle\left\langle\phi_7\right|+\rm
H.c.).
\end{eqnarray}
When we choose the initial state $|\phi_1\rangle =
|0\rangle_A|g\rangle_B|0\rangle_{AC}|0\rangle_{BC}|0\rangle_f$,
 the Hamiltonian $H_{\rm total}$ reduces to
\begin{eqnarray}\label{16}
H_{\rm
eff}&=&\Omega_{A1}(t)|\psi_1\rangle\left\langle\phi_1\right|+\Omega_{B1}(t)|\psi_1\rangle\left\langle\phi_7\right|+\rm
H.c.,
\end{eqnarray}
where $\Omega_{A1}(t)=\frac{1}{\varepsilon}\eta\Omega_{A}(t)$ and
$\Omega_{B1}(t)=\frac{1}{\varepsilon}\eta\Omega_{B}(t)$.

In order to construct the shortcuts for generating
three-dimensional entanglement by the dynamics of invariant based
inverse engineering, we need to find out the Hermitian invariant
operator $I(t)$, which satisfies
 $i\hbar \frac{\partial I(t)}{\partial t}=[H_{\rm eff}(t),I(t)]$. Since $H_{\rm eff}(t)$ possesses SU(2)
dynamical symmetry, $I(t)$ can be easily given by
\cite{MA2001,XCE2011}
\begin{eqnarray}\label{17}
I(t)=\chi(\cos\nu\sin\beta|\psi_1\rangle\langle\phi_1|+\cos\nu\cos\beta|\psi_1\rangle\langle\phi_7|+i\sin\nu|\phi_7\rangle\langle\phi_1|+\rm
H.c. ),
\end{eqnarray}
where $\chi$ is an arbitrary constant with units of frequency to
keep $I(t)$ with dimensions of energy, $\nu$ and $\beta$ are
time-dependent auxiliary parameters which satisfy the equations
\begin{eqnarray}\label{18}
\dot{\nu}&=&\Omega_{A1}(t)\cos\beta-\Omega_{B1}(t)\sin\beta,\nonumber\\
\dot{\beta}&=&\Omega\tan\nu[\Omega_{A1}(t)\cos\beta+\Omega_{B1}(t)\sin\beta].
\end{eqnarray}
Then we can derive the expressions of $\Omega_{A1}(t)$ and
$\Omega_{B1}(t)$ easily as follows:
\begin{eqnarray}\label{19}
\Omega_{A1}(t)&=&(\dot{\beta}\cot\nu\sin\beta+\dot{\nu}\cos\beta),\nonumber\\
\Omega_{B1}(t)&=&(\dot{\beta}\cot\nu\cos\beta-\dot{\nu}\sin\beta).
\end{eqnarray}
The solution of Shr\"{o}dinger equation
$i\hbar\partial|\Psi(t)\rangle/\partial t=H_{\rm
eff}(t)|\Psi(t)\rangle$ with respect to the instantaneous
eigenstates of $I(t)$ can be written as
$|\Psi(t)\rangle=\sum_{n=0,\pm}C_ne^{i\theta_n}|\Phi_n(t)\rangle$,
where $\theta_n(t)$ is the  Lewis-Riesenfeld phase in Eq.
(\ref{3}), $C_n=\langle\Phi_n(0)|\phi_1'\rangle$, and
$|\Phi_n(t)\rangle$ is the eigenstate of the invariant $I(t)$
\begin{eqnarray}\label{20}
|\Phi_0(t)\rangle&=&\cos\nu\cos\beta|\phi_1\rangle-i\sin\nu|\psi_1\rangle-\cos\nu\sin\beta|\phi_7\rangle,\nonumber\\
|\Phi_\pm(t)\rangle&=&\frac{1}{\sqrt{2}}[(\sin\nu\cos\beta\pm
i\sin\beta)|\phi_1\rangle+i\cos\nu|\psi_1\rangle\cr
&&-(\sin\nu\sin\beta\mp i\cos\beta)|\phi_7\rangle].
\end{eqnarray}
In order to transfer the population from state $|\phi_1\rangle$ to
$-|\phi_3'\rangle$, we choose the parameters as
\begin{eqnarray}\label{21}
\nu(t)=\epsilon,~~~~~~~~~\beta(t)=\frac{\pi t}{2t_f},
\end{eqnarray}
where $\epsilon$ is a time-independent small value and $t_f$ is
the total pulse duration. After the precise calculation, we can
easily obtain
\begin{eqnarray}\label{22}
\Omega_{A1}(t)&=&\frac{\pi}{2t_f}\cot\epsilon\sin\frac{\pi t}{2t_f},\nonumber\\
\Omega_{B1}(t)&=&\frac{\pi}{2t_f}\cot\epsilon\cos\frac{\pi
t}{2t_f},
\end{eqnarray}
and
\begin{eqnarray}\label{23}
\Omega_{A}(t)&=&\frac{\sqrt{g^{2}+2\eta^{2}}\pi}{2t_f}\cot\epsilon\sin\frac{\pi t}{\eta2t_f},\nonumber\\
\Omega_{B}(t)&=&\frac{\sqrt{g^{2}+2\eta^{2}}\pi}{\eta2t_f}\cot\epsilon\cos\frac{\pi
t}{2t_f}.
\end{eqnarray}
When $t=t_f$,
\begin{eqnarray}\label{24}
|\Psi(t_f)\rangle&=&-i\sin\epsilon\sin\theta|\phi_1\rangle+(-i\sin\epsilon\cos\epsilon+i\sin\epsilon\cos\epsilon\cos\theta)|\psi_1\rangle\cr
&&+(-\cos^2\epsilon-\sin^2\epsilon\cos\theta)|\phi_7\rangle,
\end{eqnarray}
where $\theta=\pi/(2\sin\epsilon)=|\theta_\pm|$ ($\theta_\pm$ are
the Lewis-Riesenfeld phases). We choose
$\theta=2N\pi(N=1,2,3...)$, then
$|\Psi(t_f)\rangle=-|\phi_7\rangle$.

On the other hand, for the initial state $|\phi_1'\rangle =
|1\rangle_A|g\rangle_B|0\rangle_{AC}|0\rangle_{BC}|0\rangle_f$,
the whole system evolves in the subspace spanned by
\begin{eqnarray}\label{25}
|\phi_1'\rangle&=&|1\rangle_A|g\rangle_B|0\rangle_{AC}|0\rangle_{BC}|0\rangle_f,\nonumber\\
|\phi_2'\rangle&=&|e_L\rangle_A|g\rangle_B|0\rangle_{AC}|0\rangle_{BC}|0\rangle_f,\nonumber\\
|\phi_3'\rangle&=&|L\rangle_A|g\rangle_B|1_L\rangle_{AC}|0\rangle_{BC}|0\rangle_f,\nonumber\\
|\phi_4'\rangle&=&|L\rangle_A|g\rangle_B|0\rangle_{AC}|0\rangle_{BC}|1_L\rangle_f,\nonumber\\
|\phi_5'\rangle&=&|L\rangle_A|g\rangle_B|0\rangle_{AC}|1_L\rangle_{BC}|0\rangle_f,\nonumber\\
|\phi_6'\rangle&=&|L\rangle_A|e_L\rangle_B|0\rangle_{AC}|0\rangle_{BC}|0\rangle_f,\nonumber\\
|\phi_7'\rangle&=&|L\rangle_A|L\rangle_B|0\rangle_{AC}|0\rangle_{BC}|0\rangle_f.
\end{eqnarray}
The effective Hamiltonian in the subspace is
\begin{eqnarray}\label{26}
H_{\rm
eff}&=&\Omega_{A1}(t)|\psi_1'\rangle\left\langle\phi_1'\right|+\Omega_{B1}(t)|\psi_1'\rangle\left\langle\phi_7'\right|+\rm
H.c.,
\end{eqnarray}
where
$|\Psi_{1}'\rangle=\frac{1}{\epsilon}(\eta|\phi_{2}'\rangle-g|\phi_{4}'\rangle+\eta|\phi_{6}'\rangle)$.

With the same way as above, we can realize the transition from
$|\phi_{1}'\rangle$ to $|\phi_{7}'\rangle$.

Then we make one qubit operation on atom A to make $|g\rangle_{A}$
become $-|g_{A}\rangle$ with the help of laser pulses resonant
with A atomic transition
$|g\rangle_{A}\leftrightarrow|e_{R}\rangle_{A}$ and
$|R\rangle_{A}\leftrightarrow|e_{R}\rangle_{A}$ with the
corresponding Rabi frequencies $\Omega_{g}(t)$ and
$\Omega_{R}(t)$. In this step, the Hamiltonian in the interaction
picture can be written as ($\hbar=1$)
\begin{eqnarray}\label{27}
H_{2}=\Omega_{g}(t)|e_{R}\rangle_{A}\langle
g|+\Omega_{R}(t)|e_{R}\rangle_{A}\langle R|+\rm{H.c.}
\end{eqnarray}

With the same method as above, we can choose
\begin{eqnarray}\label{28}
\Omega_{g}(t)=\frac{\pi}{2t_{f}}\cot\epsilon\sin\frac{\pi
t}{2t_{f}},\nonumber\\
\Omega_{R}(t)=\frac{\pi}{2t_{f}}\cot\epsilon\cos\frac{\pi
t}{2t_{f}}.
\end{eqnarray}
Here we choose $t=2t_{f}$, and with the similar processes as above
we can realize the transformation from $|g_{A}\rangle$ to
$-|g_{A}\rangle$.

Up to now, the initial state
\begin{eqnarray}\label{29}
|\Psi(0)=\frac{1}{3}(|0\rangle_{A}+|1\rangle_{A}
+|g\rangle_{A})|g\rangle_{B}|0\rangle_{AC}|0\rangle_{BC}|0\rangle_{f}
\end{eqnarray}
of the whole system has evolved into the state
\begin{eqnarray}\label{30}
|\Psi\rangle&=&\frac{1}{\sqrt{3}}(|R\rangle_A|R\rangle_B+|L\rangle_A|L\rangle_B
+|g\rangle_A|g\rangle_B)|0\rangle_{AC}|0\rangle_{BC}|0\rangle_{f}.
\end{eqnarray}
Ignoring the global phase, the two atoms are in three-dimensional
entanglement, with the cavity-modes and the fiber mode in vacuum
state.

\noindent {\bf 4. Numerical simulations and feasibility analysis}

\begin{figure}[ht]\centering
\scalebox{0.9}{\includegraphics{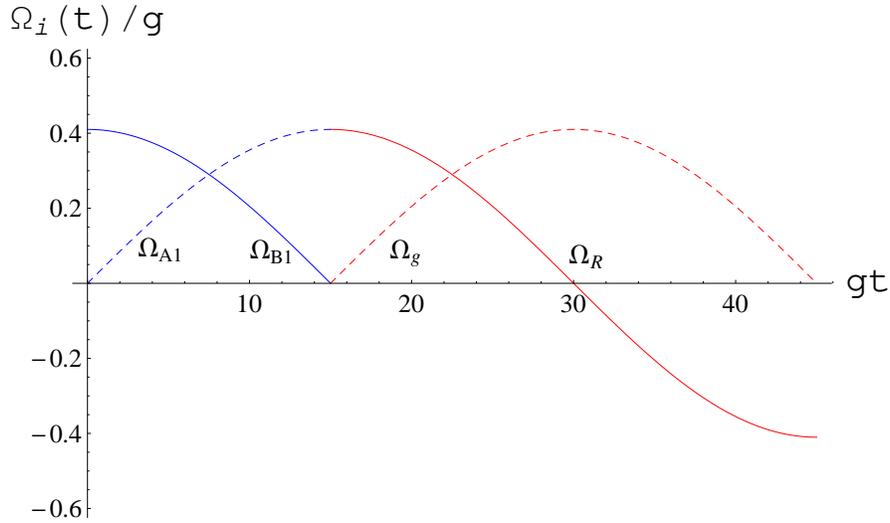}}\caption{Temporal
profile of the time dependence Rabi frequencies $\Omega_{i}(t)/g$
versus $gt$ with $\Omega_{A1}(t)$ (dash blue line),
$\Omega_{B1}(t)$ (solid blue line), $\Omega_{g}(t)$ (dash red
line), $\Omega_{R}(t)$ (solid red line).}\label{fig3}
\end{figure}

\begin{figure}[ht]\centering
\label{fig4a}
\includegraphics[width=3.2in,height=2.8in]{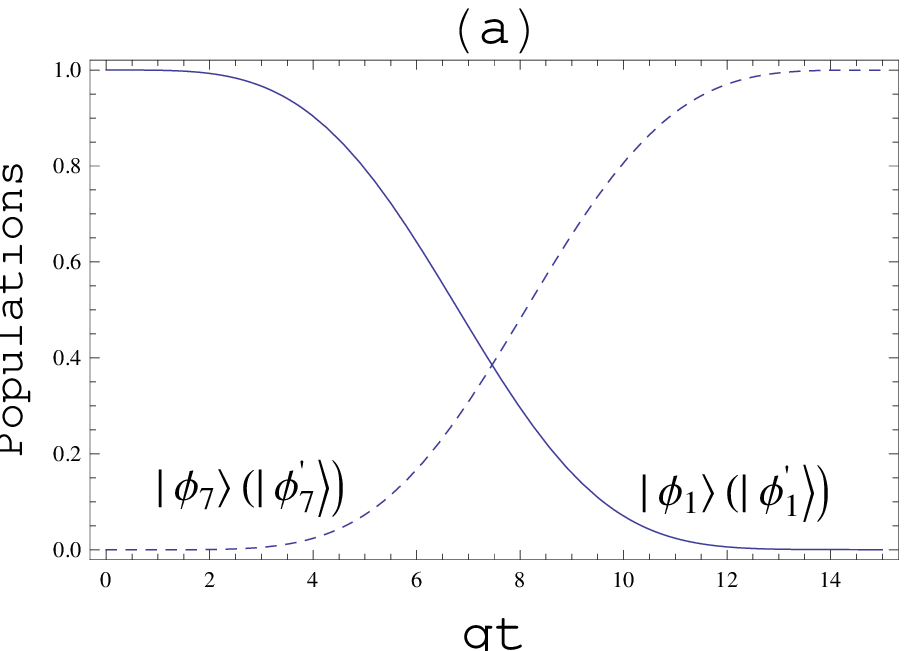}
\label{fig4b}
\includegraphics[width=3.2in,height=2.8in]{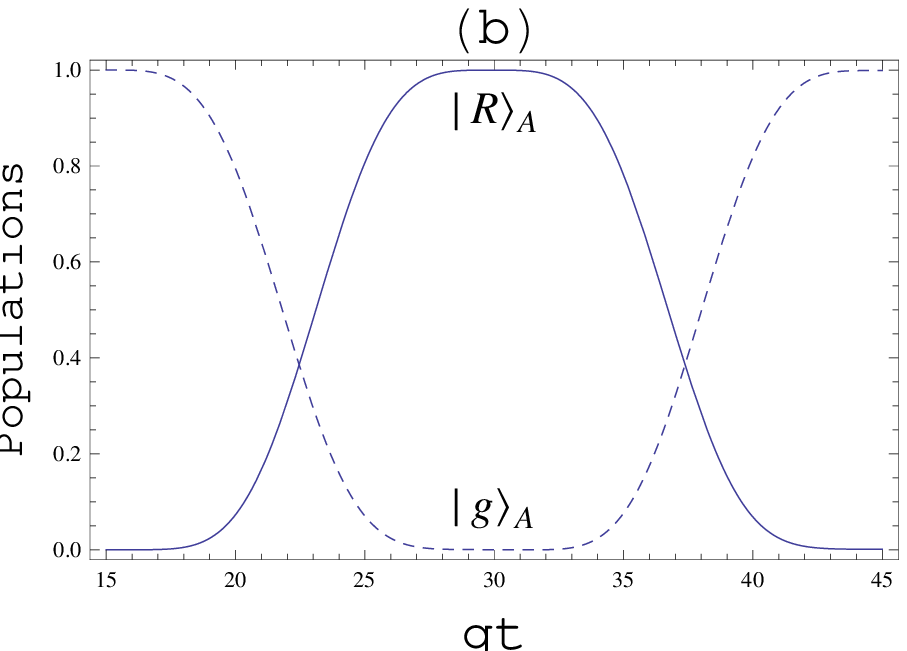}
\caption{(a) Time evolutions of the populations of the
corresponding system states with the initial states
$|\phi_1\rangle (|\phi_{1}'\rangle)$. (b) Time evolutions of the
populations with the initial state $|g\rangle_{A}$. The system
parameters are set to be $\epsilon= 0.25$, $g_{\rm A}=g_{\rm B}=g$
with $ t_{f}=15/g.$.}\label{fig4}
\end{figure}

In the following, we present the numerical validation of the
mechanism proposed for the generation of three-dimensional
entanglement of the two atoms. Fig. \ref{fig3} shows the
time-dependence laser pulse $\Omega_{i}(t)/g$ as a function of
$gt$ for a fixed value $\epsilon= 0.25$, and $t_{f}=15/g$. With
these parameters the Zeno condition can be met well. The
populations of states $|\phi_{1}\rangle (|\phi_{1}'\rangle)$ and
$|\phi_{7}\rangle (|\phi_{7}'\rangle)$ swap perfectly when
$t=t_{f}$, as shown in Fig. \ref{fig4}(a), and the populations of
states $|R\rangle_{A}$ and $|g\rangle_{A}$ also swap perfectly
when $t=2t_{f}$ as shown in Fig. \ref{fig4}(b).

\begin{figure}\centering
\scalebox{0.9}{\includegraphics{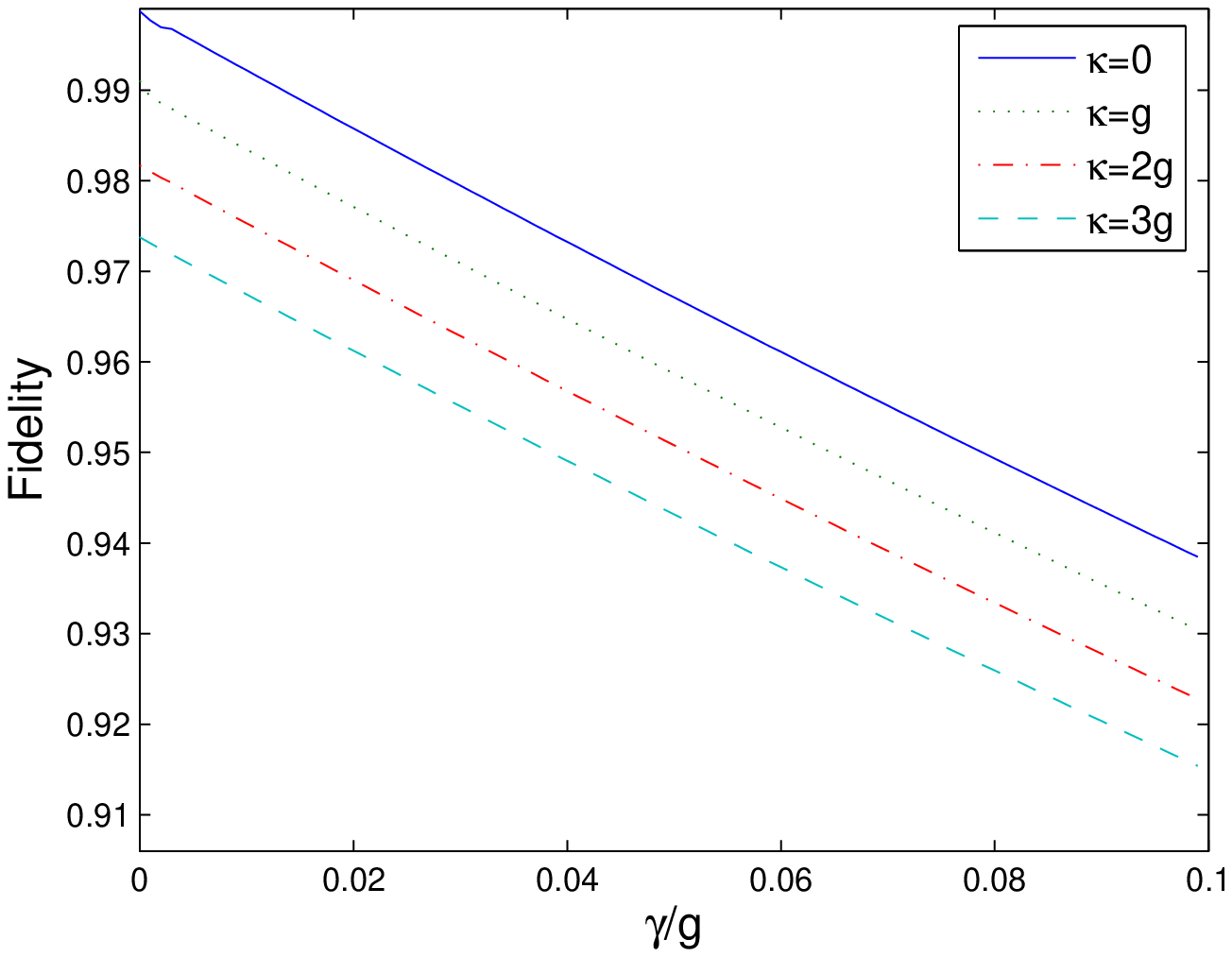}}\caption{The effect of
atomic spontaneous emission $\gamma$ on the fidelity of the
three-dimensional entanglement with different values of the photon
leakage rates $\kappa$ of cavities or fiber.}\label{fig5}
\end{figure}

In addition, whether a scheme is available largely depends on the
robustness to the loss and decoherence. so we consider the effects
of loss and decoherence on the entanglement generation. The
corresponding master equation for the whole system density matrix
$\rho(t)$ has the following form:
\begin{eqnarray}\label{31}
\dot{\rho(t)}&=&-i[H,\rho(t)]-\sum_{j=L,R}\frac{\kappa_{j}^{f}}{2}[b_{j}^{\dag}b_{j}\rho(t)-2b_{j}\rho(t)b_{j}^{\dag}+\rho(t)b^\dag
b ]\cr
&&-\sum_{j=L,R}\sum_{i=A,B}\frac{\kappa_{j}^{i}}{2}[a_{ij}^{\dag}a_{ij}\rho(t)-2a_{ij}\rho(t)a_{ij}^{\dag}+\rho(t)a_{ij}^\dag
a_{ij} ]\cr
&&-\sum_{j=A,B}\sum_{h=0,1,L,R}\frac{\gamma_{jh}^{A}}{2}[\sigma^{A}_{e_j,e_j}\rho(t)-2\sigma^{A}_{h,e_j}\rho(t)\sigma^{A}_{e_j,h}+\rho(t)\sigma^{A}_{e_j,e_j}]
\cr
&&-\sum_{j=A,B}\sum_{m=g,L,R}\frac{\gamma_{jm}^{B}}{2}[\sigma^{B}_{e_j,e_j}\rho(t)-2\sigma^{B}_{m,e_j}\rho(t)\sigma^{B}_{e_j,m}+\rho(t)\sigma^{B}_{e_j,e_j}],
\end{eqnarray}
where $H=H_{1}+H_{2}$. $\kappa_{j}^{f}$ is the photon leakage rate
of $j$th fiber mode, $\kappa_{j}^{i}$ is the photon leakage rate
of $j$-circular polarization mode in $i$th cavity,
$\gamma_{jh(jm)}^{A(B)}$ is $j$th atomic spontaneous emission rate
of cavity $A(B)$ from the excited state $|e\rangle_{j}$ to the
corresponding ground state $|h(m)\rangle$.
$\sigma_{e_j,e_j}=|e_j\rangle\left\langle e_j\right| (j=A,B)$,
$\sigma_{e_j,h(h,e_j)}=|e_j(h)\rangle\left\langle h(e_j)\right|$
and $\sigma_{e_j,m(m,e_j)}=|e_j(m)\rangle\left\langle
m(e_j)\right|), (j=A,B)$. For simplicity, we assume
$\kappa_{j}^{f}=\kappa_{j}^{i}=\kappa$,
$\gamma_{jh(jm)}^{A(B)}=\gamma$. The initial condition $\rho(0)=
|\Psi_0\rangle\left\langle \Psi_0\right|$. Fig. \ref{fig5} shows
the fidelity $F = \langle\Psi_0|\Psi(t)\rangle$ as a function of
the dimensional parameter $\gamma/g$ with different values of
$\kappa$ by numerically solving the master equation (\ref{31}).
From Fig. \ref{fig5} we can see that, the fidelity for
three-dimensional entanglement is higher than $93\%$ when
$\gamma=0.1g$ and $\kappa=g$. It shows that our scheme is robust
against decoherence caused by photon leakage of cavities and
fiber, and atomic spontaneous emission.

Now we give a brief analysis of the feasibility in experiment of
our scheme. The appropriate atomic level configuration can be
obtained from the hyperfine structure of cold alkali-metal atoms
\cite{TWS2007,BWH2009,MLM2011}. Here we adopt the ${}^{133}\!$Cs.
5S$_{1/2}$ ground level $|F=3,m=2\rangle(|F=3,m=-2\rangle)$
corresponds to $|R\rangle(|L\rangle)$ and
$|F=2,m=1\rangle(|F=2,m=-1\rangle)$ corresponds to
$|0\rangle(|1\rangle)$, respectively, while 5P$_{3/2}$ excited
level $|F=3,m=1\rangle(|F=3,m=-1\rangle)$ corresponds to
$|e_R\rangle(|e_L\rangle)$. Other hyperfine levels in the
ground-state manifold can be used as $|g\rangle$ for atom A. For
atom B, the states $|R\rangle,|L\rangle$ and $|g\rangle$
correspond to $|F=2,m=-1\rangle,|F=2,m=1\rangle$ and
$|F=3,m=0\rangle$ of 5S$_{1/2}$ ground levels, respectively. And
$|e_R\rangle(|e_L\rangle)$ corresponds to
$|F=3,m=-1\rangle(|F=3,m=1\rangle)$ of 5P$_{3/2}$ excited level.

\noindent{\bf 5. Conclusion}

In conclusion, we have proposed a scheme for generating
three-dimensional entanglement of two spatially separated atoms
through the shortcut to adiabatic passage and QZD. We also study
the influences of system parameters, such as photon leakage of
cavities and fiber, and atomic spontaneous emission, on the
fidelity through numerical simulation. The numerical simulation
results show that our scheme is very robust against the system
parameters.

\begin{center}
{\bf{Acknowledgment}}
\end{center}

This work was supported by the National Natural Science Foundation
of China under Grant Nos. 11464046 and 61465013.

\end{document}